\documentclass[acmtog,nonacm]{acmart}

\usepackage{booktabs}
\usepackage{natbib}
\usepackage{xspace}
\usepackage[nolist]{acronym}
\usepackage{bm}

\usepackage{siunitx}
\robustify\fontseries
\robustify\selectfont
\newcommand{\winner}[1]{\fontseries{b}\selectfont{#1}}

\usepackage{pifont}
\newcommand{\cmark}{\checkmark}%
\newcommand{\xmark}{\scalebox{0.85}{\ding{53}}}%
\usepackage{adjustbox}
\newcolumntype{R}{%
    >{\adjustbox{angle=90}\bgroup}%
    l%
    <{\egroup}%
}
\newcommand*\rot{\multicolumn{1}{R}}%

\usepackage{soul}
\usepackage{amsmath}
\usepackage{microtype}
\usepackage{wrapfig}

\def\figurePath{figures_arxiv/}

\def\myfigure#1#2{\begin{figure}[htb]\centering\includegraphics*[width = \linewidth]{\figurePath#1}\caption{#2}\label{fig:#1}\end{figure}}

\def\mycfigure#1#2{\begin{figure*}[t]\centering\includegraphics*[clip, width = \linewidth]{\figurePath#1}\caption{#2}\label{fig:#1}\end{figure*}}

\newcommand{\new}[1]{#1}

\newcommand{\eg}{e.g., }
\newcommand{\ie}{i.e., }

\newcommand{\refSec}[1]{Sec.~\ref{sec:#1}}
\newcommand{\refFig}[1]{Fig.~\ref{fig:#1}}
\newcommand{\refEq}[1]{Eq.~\ref{eq:#1}}
\newcommand{\refTbl}[1]{Tbl.~\ref{tbl:#1}}
\newcommand{\refFigStart}[1]{Figure~\ref{fig:#1}}

\newcommand{\unsure}[1]{{\sethlcolor{yellow}\hl{#1}}}

\soulregister\ref7
\soulregister\cite7
\soulregister\refFig7
\soulregister\refFigStart7
\soulregister\cite7
\soulregister\ref7
\soulregister\pageref7
\soulregister\shortcite7
\soulregister\namecite7
\soulregister\eg7
\soulregister\ie7
\soulregister\etal7
\soulregister\reftbl7
\soulregister\name7
\soulregister\unsure7

\DeclareGraphicsExtensions{.png,.jpg,.pdf,.ai,.psd}
\newcommand{\mysection}[2]{\section{#1}\label{sec:#2}}
\newcommand{\mysubsection}[2]{\subsection{#1}\label{sec:#2}}

\begin{acronym}
\acro{2AFC}{Two-alternative Forced Choice}
\acro{AdaIN}{Adaptive Instance Normalization}
\acro{BRDF}{Bi-directional Reflectance Distribution Function}
\acro{CNN}{Convolutional Neural Network}
\acro{svBRDF}{Spatially-varying Bi-directional Reflectance Distribution Function}
\acro{VAE}{Variational Auto-encoder}
\end{acronym}

\newcommand{\mymath}[2]{\newcommand{#1}{\TextOrMath{$#2$\xspace}{#2}}}

\mymath{\location}{\mathbf x}
\mymath{\scale}{s}
\mymath{\flashImage}{L}
\mymath{\outgoingDirection}{\omega_\mathrm o}
\mymath{\render}{\mathbf R}
\mymath{\materialModel}{f}
\mymath{\materialModelParameters}{\theta}
\mymath{\lossOld}{\mathcal L'}
\mymath{\textureComparisonOld}{\mathcal T'}
\mymath{\patchOld}{\mathcal P'}
\mymath{\loss}{\mathcal L}
\mymath{\textureComparison}{\mathcal T}
\mymath{\patch}{\mathcal P}
\mymath{\stationarity}{\mathcal S}
\mymath{\materialCode}{\mathbf z}
\mymath{\flashImageEncoder}{g}
\mymath{\expectation}{\mathbb E}
\mymath{\latentDimension}{n_\mathrm z}
\mymath{\spectralWeight}{\lambda}
\mymath{\Zdimension}{64}
\mymath{\fineFlashImage}{\flashImage^\star}
\mymath{\fineMaterialModelParameters}{\materialModelParameters^\star}
\mymath{\fineMaterialCode}{\materialCode^\star}

\setcopyright{none}

\setcitestyle{square}

\setcopyright{acmlicensed}\acmJournal{TOG}
\acmYear{2021}\acmVolume{40}\acmNumber{6}\acmArticle{1}\acmMonth{12} \acmDOI{10.1145/3478513.3480507}

\begin{document}

\title{Generative Modelling of BRDF Textures from Flash Images}

\author{Philipp Henzler}
\affiliation{
  \department{Department of Computer Science}
  \institution{University College London}
    \country{United Kingdom}
  }
\email{p.henzler@cs.ucl.ac.uk}

\author{Valentin Deschaintre}
\affiliation{%
  \institution{Adobe Research and Imperial College London }
  \country{United Kingdom}
  }
  

\author{Niloy J. Mitra}
\affiliation{%
  \institution{University College London and Adobe Research}
    \country{United Kingdom}
  }

\author{Tobias Ritschel}
\affiliation{
  \department{Department of Computer Science}
  \institution{University College London}
    \country{United Kingdom}
  }

\renewcommand{\shortauthors}{Henzler et al.}

%

\begin{abstract}
We learn a latent space for easy capture, consistent interpolation, and efficient reproduction of visual material appearance.
When users provide a photo of a stationary natural material captured under flashlight illumination, first it is converted into a latent material code.
Then, in the second step, conditioned on the material code, our method produces an infinite and diverse spatial field of BRDF model parameters (diffuse albedo, normals, roughness, specular albedo) that subsequently allows rendering in complex scenes and illuminations, matching the appearance of the input photograph.
Technically, we jointly embed all flash images into a latent space using a convolutional encoder, and --conditioned on these latent codes-- convert random spatial fields into fields of BRDF parameters using a convolutional neural network (CNN).
We condition these BRDF parameters to match the visual characteristics (statistics and spectra of visual features) of the input under matching light.
A user study compares our approach favorably to previous work, even those with access to BRDF supervision. 
\noindent
Project webpage: \url{https://henzler.github.io/publication/neuralmaterial/}.

\end{abstract}



\begin{teaserfigure}
    \centering
    \includegraphics[width=\linewidth]{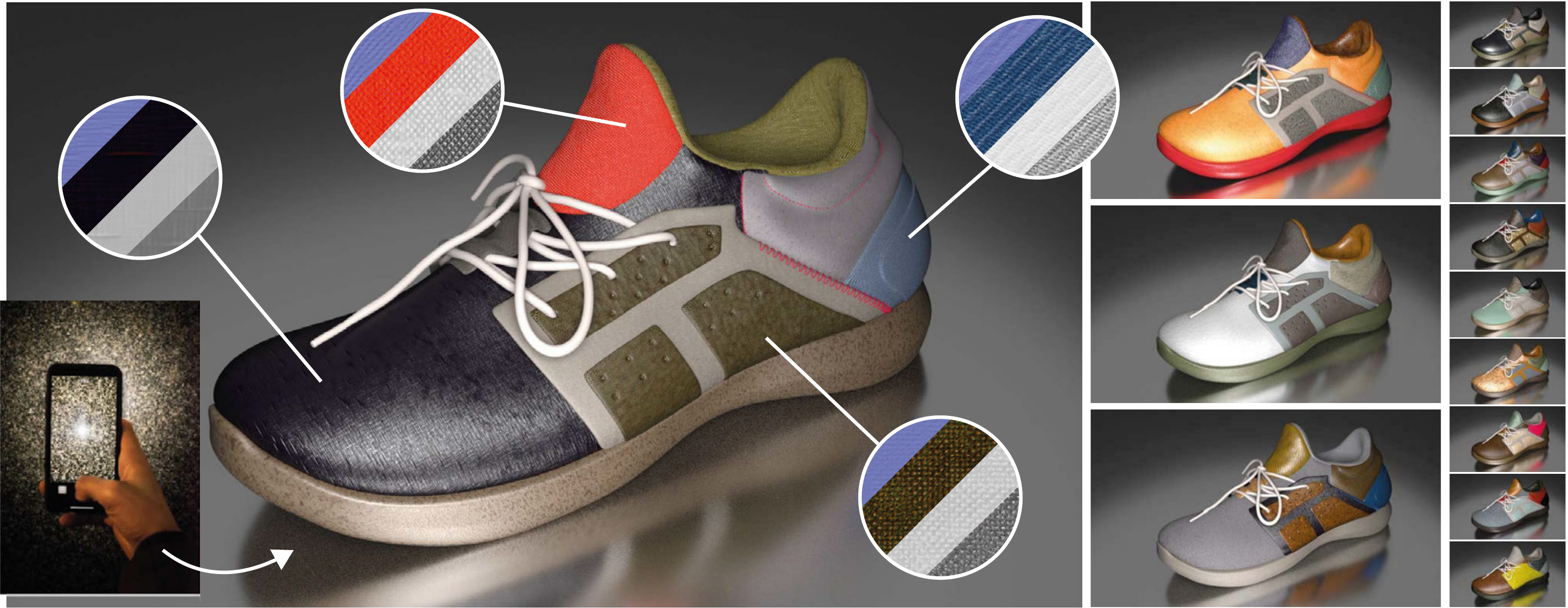}
    \caption{
        Results from our generative model of BRDF maps, assigned to a 3D object of a shoe.
        Circular insets show the diffuse, normal, roughness and specular maps.
        Our model outputs a space of BRDF materials that can be sampled from, projected to, and interpolated across.
        The BRDF generative model is trained exclusively from a set of ($306$) RGB flash images (example shown in the lower left inset) without any BRDF supervision, and shortly fine-tuned in the case of material capture to best match the input picture.
    }
    \label{fig:Teaser}
\end{teaserfigure}

\maketitle

\section{Introduction}
Rendering realistic images for feature films or computer games requires adequate simulation of light transport.
Besides geometry and illumination, an important factor is material appearance.

Material appearance has three aspects of variation:
First, when view or light direction change, reflected light changes.
The physics of this process is well-understood and can be simulated provided the input parameters are available.
Second, behavior changes across materials.
For example, leather reacts differently to light or view changes than paper would, yet, different forms of leather clearly share visual properties, \ie form a (material) space.
Third, appearance details depend on spatial position.
Different locations in the same leather exemplar behave differently but share the same visual statistics \cite{portilla2000parametric}, \ie they form a \emph{texture}.

Classic computer graphics captures appearance by \emph{reflection models}, which predict for a given i)~light-view configuration, ii)~material, and iii)~spatial position, how much light is reflected.
Typically, the first variation (light and view direction) is covered by \emph{BRDF models}, analytic expressions, such as \citet{phong1975illumination} which map the light and view direction vector to scalar reflectance.
The second variation (material) is covered by choosing \emph{BRDF model parameters}, such as \new{specularity or roughness}.
In practice, it can be difficult, given a desired appearance, to choose those parameters \eg how to make a leather look more like the one on a nice jacket.
One can measure BRDF model parameters, but it traditionally requires complex capture hardware for accurate results.
The third variation (spatial) is addressed by storing multiple BRDF model parameters in images of finite size --often referred to as \ac{svBRDF} maps-- or writing functional expressions to reproduce their behaviour.
It is even more challenging to choose these parameters to produce something coherent like leather, in particular over a large to very large spatial extent.
Additionally, storing all these values requires substantial memory and programming functional expressions to mimic their statistics requires expert skills and time.
Capturing the spatial variation of BRDF model parameters over space using sensors requires even more complex hardware~\cite{schwartz-2013-dome2wmam}.

Addressing those issues, we provide a reflectance model to jointly generalize across all of these three axes.
Instead of using analytic parameters, we parametrize appearance by latent codes from a learned space and our decoder weights, allowing for acquisition, interpolation and generation.
Without involved capture equipment, these codes are produced by presenting the system a simple 2D flash image, which is then embedded into the latent space.
Avoiding to store any finite image texture, we learn a second mapping to produce \ac{svBRDF} maps from the infinite random field (noise) on-the-fly, conditioned on the latent material code and decoder weights. 
Instead of using any advanced capture device for learning, flash images will be the only supervision we use.
This unsupervised approach allows us to consider our decoder weights as part of the latent representation, which we fine-tune at test time in a few minutes.
\myfigure{Space}{
\textbf{BRDF space}. 
From a flash image, which contains sparse observations across material, space and view-light \textbf{(left)}
we map to a latent code $\materialCode$ / $\fineMaterialModelParameters$ \textbf{(middle)}
so that changes in these code can be decoded to enable \textbf{(right)} material synthesis (holding material fixed and moving spatially), material morphing (holding space and view/light fixed and changing material), or classical shading and material generation (points in the latent space).}

A use case of our approach is shown in \refFig{Teaser}.
First, a user provides a ``flash image'', a photo of a flat material sample under flash illumination.
This sample is embedded as a code into a latent space using a CNN and used to fine-tune our decoders weights.
This code and weights can then be manipulated, \eg interpolated with a different material. 
Conditioned on this code, our fine-tuned decoder can generate an infinite field of BRDF to be directly used in rendering. 

For training, we solely rely on real flash images.
The key insight, inspired by \citet{aittala2016reflectance}, is that these flash images reveal the same material at different image locations --they are stationary-- but under different view and light angles.
Using this constraint, \citet{aittala2016reflectance} were able to decompose a small patch of a single input image to capture the parameters of a material model that could then be rendered under novel view or light directions.
However, this covers only part of the generalization we are targeting: it generalizes across view and light, but not across location or material. 
Further, they perform an optimization for every exemplar, requiring time in the order of an hour, while ours takes minutes only.

In summary, our main contributions are
\begin{itemize}
\item a generative model of a BRDF material texture space;
\item generation of maps that are diverse over the infinite plane;
\item a flash image dataset of materials enabling our training with no BRDF parameter supervision or synthetic data
\end{itemize}

\noindent Our implementation will be publicly available upon acceptance.

\mysection{Previous Work}{PreviousWork}
Our work has background in texture analysis, appearance modelling and design spaces as summarized in \refTbl{PreviousWork} and discussed next.

\begin{table}[h]
    \setlength{\tabcolsep}{3.0pt}
    \centering
    \caption{Comparison of features between different previous methods. We distinguish methods producing RGB from those generating \textbf{BRDF} or \textbf{\ac{svBRDF}}, whether those can be \textbf{Non-Stationary} and \textbf{Infinitely} sampled.
    We also distinguish if their  results for one input can be \textbf{Diverse}, if they form a \textbf{Space} which can be queried and how \textbf{Fast} direct sampling is.}
    \label{tbl:PreviousWork}
    \vspace{-0.3cm}
    \begin{tabular}{rlccccccc}
        \multicolumn1c{Method}&
        \multicolumn1c{Supervision}&
        \rot{BRDF}&
        \rot{\ac{svBRDF}}&
        \rot{Non-Stat.}&
        \rot{Infinite}&
        \rot{Diverse}&
        \rot{Space}&
        \rot{Fast Gen.}\\
        \toprule
        Classic texture synth&
        RGB&
        \xmark&
        \xmark&	
        \cmark&
        \cmark&	
        \cmark&	
        \xmark&
        \cmark\\
        \citet{matusik2005texture}&
        RGB&
        \xmark&
        \xmark&
        \cmark&
        \cmark&
        \cmark&
        \cmark&
        \xmark\\
        \citet{matusik2003data}&
        BRDF&
        \cmark&	
        \xmark&	
        \xmark&	
        \xmark&	
        \cmark&	
        \cmark&
        \cmark\\
        \citet{georgoulis2017reflectance}&
        BRDF&
        \cmark&	
        \xmark&	
        \xmark&	
        \xmark&	
        \xmark&
        \cmark&
        \cmark\\
        \citet{deschaintre2018single}&
        \ac{svBRDF}&
        \cmark&
        \cmark&
        \xmark&
        \xmark&
        \xmark&
        \xmark&
        \cmark\\
        \citet{zhao2020joint}&
        Flash image&
        \cmark&
        \cmark&
        \cmark&
        \xmark&
        \xmark&
        \xmark&
        \xmark\\
        \citet{aittala2016reflectance}&
        Flash image&
        \cmark&
        \cmark&
        \cmark&
        \xmark&
        \xmark&
        \xmark&
        \xmark\\
        \citet{gao2019deep}&	
        \ac{svBRDF}&
        \cmark&
        \cmark&
        \cmark&
        \xmark&
        \xmark&
        \xmark&
        \xmark\\
        \citet{guo2020materialgan}&	
        \ac{svBRDF}&
        \cmark&
        \cmark&
        \cmark&
        \xmark&
        \xmark&
        \cmark&
        \xmark\\
        Ours&
        Flash image&
        \cmark&
        \cmark&
        \xmark&
        \cmark&
        \cmark&
        \cmark&
        \cmark
        \\
        \bottomrule
    \end{tabular}
\end{table}

\subsection{Textures in Graphics}
A classic definition of texture is defined by \citet{julesz1965texture}: \textit{a texture is an image full of features that in some representation have the same statistics.}
\citet{portilla2000parametric} have provided a practical method to compute representations in which to do statistics on, using linear filters on multiple scales.

\citet{perlin1985noise} was first to capture the fractal \cite{mandelbrot1983fractal} stochastic variation of appearance in a model applicable to Computer Graphics.
His approach is simple --a linear combination of noise at different scales-- yet extremely powerful, and has led to extensive use in computer games and production rendering.
Wavelet noise \cite{cook2005wavelet} moved this idea further by band-limiting the noise that is combined.
Such methods can be used for materials, \eg gloss maps, bump maps, etc.
It however does not provide a solution to acquire a texture from an exemplar, which is left to manual adjustment.

To generate textures from exemplars, non-parametric sampling \cite{efros1999texture}, vector quantization \cite{wei2000fast}, optimization \cite{kwatra2005texture} or nearest-neighbour field synthesis (PatchMatch \cite{barnes2009patchmatch}) have been proposed. They however have issues in computational scalability and lack intuitive control, limiting their adoption in production rendering or games.

The word ``texture'' can be ambiguous to mean stochastic variation, as well as images attached on surfaces to localize color features.
Here, we focus on stochastic variation in the sense of \citet{julesz1965texture} or \citet{portilla2000parametric}.

Our approach is inspired by deep learning-based texture synthesis 
\cite{simonyan2014very,gatys2015texture,sendik2017deep,ulyanov2016texture,johnson2016perceptual,ulyanov2017improved,shaham2019singan,bergmann2017learning,zhou2018non,karras2019style} which ideas we extend and apply to BRDFs.
We detail their background in \refSec{TextureSynthesis}

\subsection{Material Modeling}
Representing appearance in simulation-based graphics has been an active research field for decades.
The survey by \citet{guarnera2016brdf} presents detailed discussion of the many different material model and BRDF acquisition approach.
In our method, we use a state-of-the-art micro-facet BRDF model \cite{cook1982reflectance}, and focus on deep-learning based material modelling and acquisition.

Many methods have been proposed to acquire materials using data-driven approaches.
\citet{matusik2003data} proposed a data driven BRDF linear model. More recently, \citet{rematas2016deep} extract reflectance maps from 2D images using a CNN trained in a supervised manner. Materials and illuminations acquisition were further explored by \citet{georgoulis2017reflectance}.
\citet{deschaintre2018single} proposed a rendering loss to capture svRBDFs from flash images.
\citet{namPracticalSvBRDF_2018} jointly reconstructed \ac{svBRDF}, normals, and 3D geometry in an iterative inverse-rendering setup towards a practical acquisition setup, while different methods relied on deep learning to estimate object shape and \ac{svBRDF} from one or multiple images \cite{li2018learning, Boss2020-TwoShotShapeAndBrdf, Deschaintre21}. \citet{Li:Synthesizing3D:2019} propose a weakly supervised learning-based method for generating novel category-specific 3D shapes and demonstrate that it can help in learning material-class specific \acp{svBRDF} from images distributions. \citet{Wenjie:Inexact_2018} used a mixture of images and procedural material maps to train a network for modeling \acp{svBRDF}. \citet{Yiwei_inverseProcTexture:2019} developed a reduced \ac{svBRDF} model, using only diffuse and normal channels, towards solving inverse procedural textures matching from reference, while \citet{Guo:BayseianInference:2019} used Bayesian inference for material synthesis. Recently, \citet{Shi:Match:2020} developed a differentiable material graph nodes library to optimize material parameters to match an input material, given material graphs.

U-net \cite{ronneberger2015u} inspired many approaches for image to image translation to translate RGB pixels to material attributes \cite{li2018learning,li2017modeling,li2018materials, deschaintre2018single}.
Most work now includes a differentiable shading step \cite{liu2017material,li2018learning,deschaintre2018single, deschaintre2019flexible,guo2020materialgan} such as we do here.
\citet{gao2019deep} and \citet{guo2020materialgan} propose to use a post-optimization in an encoded latent space, improving an initial material estimation, and comparing renderings of their results directly to their input pictures. \citet{Deschaintre2020_fineTune} propose to fine tune their material acquisition network on \ac{svBRDF} parameter examples to transfer them to a larger scale.
\new{\citet{zhou2021adversarial} propose a partially unsupervised training approach, allowing to use additional real data without ground truth maps. With our approach, we completely remove the need for ground truth maps as it solely relies on real flash photographs.}
\new{\citet{guo2021highlight} address the issue of strong highlights baked into \ac{svBRDF} maps through highlight-aware convolutions and an attention-based feature selection module. Our design, focused on stationarity of textures, inherently prevents any flash residual to be left in the results.}

All these approaches focus on capturing a single instance of a \ac{svBRDF} map, but with little or no editing options across materials (space) or generalization across the spatial domain (diversity). For rapid materials generation, \citet{zsolnaifeher18gms} propose to use Gaussian process regression.

However, most of these methods require synthetic \ac{svBRDF} supervision for training, while we focus on directly learning from flash images without access to channel-level supervision. In particular this removes the risk of domain gap between synthetic and real materials and enables our fine-tuning approach.
 
We take inspiration from \citet{aittala2016reflectance} who extended the approach of \citet{gatys2015texture} to generate \ac{svBRDF} parameter maps from a single picture of a stationary material exemplar and propose an approach for improved diversity, generation and quality.
 
\subsection{Spaces-of}
Spaces of color \cite{nguyen2015data}, materials \new{\cite{matusik2003data,guo2020materialgan, gao2019deep}}, textures \cite{matusik2005texture}, faces \cite{blanz1999morphable}, human bodies \cite{allen2003space}, and more have been useful in graphics for content creation and edition.
\citet{matusik2005texture} has devised a space of textures.
Here, users can interpolate combinations of visually similar textures.
They warp all pairs of exemplars to each other and constructs graph edges for interpolation when there is evidence that the warping is admissible.
To blend between them, histogram adjustments are made.
Consequently, interpolation between exemplars does not take a straight path in pixel space from one to the other, but traverses only valid regions.
Photoshape \cite{park2019photoshape} learns the relation of given material textures over a database of 3D objects. \citet{serrano2016materialediting} allow users to semantically control captured BRDF data. They represent BRDFs using the derived principal component basis~\cite{matusik2003data} and map the first five PCA components to semantic attributes through learned radial basis functions. \new{Similar to our method, \citet{guo2020materialgan} and \citet{gao2019deep} produce spaces of materials that can be interpolated.}
We take inspiration from this body of work and build a
space allowing \acp{svBRDF} generation and interpolation.

\mysection{Background}{Background}

\mysubsection{Flash Images}{Flash Images}
\citet{aittala2016reflectance} leveraged the fact that a single flash image of a stationary material reveals multiple realizations of the same reflectance statistics under different light and view angles.
We will now recall a simplified definition of their approach. 

A flash image is an RGB image of a material, taken in conditions where a mobile phone's flashlight is the dominant light source.
We write $\flashImage(\location)$ to denote the RGB radiance value at every image location \location.
The illumination is \new{expected} to be an isotropic point light collocated with the camera.
Further, the geometry is assumed to be flat and captured in a fronto-parallel setting, so that the direction from light to every image location in 3D is known.
Self-occlusion and parallax are assumed to be negligible.

Reflectance is parameterized by a \emph{material}, represented as a function $\materialModel(\location)$ mapping image location \location to shading model parameters, including the shading normal.
Under these conditions, the reflected radiance is
$
\flashImage=\render\materialModel
$, where \render is the \emph{differentiable} rendering operator, mapping shading model parameters to radiance.

A material \materialModel explains a flash image \flashImage if it is \emph{visually similar} to \flashImage when rendered.
Unfortunately, without further constraints, there are many materials to explain the flash image.
This ambiguity can be resolved when assuming that the material \materialModel is \emph{stationary}.
We say a material is stationary, if local statistics of the shading model parameters $f$ do not change across the image.
  
Putting both --visual similarity and stationarity-- together, the best material from a family $\materialModel_\materialModelParameters$ of material mapping functions parameterized by a vector \materialModelParameters, can be found by minimizing a loss:
\begin{align}
\label{eq:ImageLoss}
\lossOld(\materialModelParameters):=
\textureComparison(
\flashImage,
\render\materialModel_\materialModelParameters)
+
\lambda
\stationarity(\materialModel_\materialModelParameters)
,
\end{align}
where $\textureComparison(\flashImage,\render\materialModel_\materialModelParameters)$ is a metric of visual similarity between a flash image $\flashImage$ and a differentiable rendering $\render\materialModel_\materialModelParameters$, and $\stationarity(\materialModel)$ is a measure of stationarity of a material map \materialModel.

\mycfigure
{Architecture}
{\textbf{Our architecture}.
Starting from an exemplar \textbf{(top-left)} our trained encoder encodes the image to a compact latent space variable $z$.
Additionally, a random infinite field is cropped with the same spatial dimensions as the flash input image. The noise crop is then reshaped based on a convolutional U-Net architecture. Each convolution in the network is followed by an Adaptive Instance Normalization (AdaIN) layer \cite{huang2017arbitrary} reshaping the statistics (mean $\mu$ and standard deviation $\sigma$) of features.
A learned affine transformation $T$-s per layer maps $z$ to the desired $\mu$-s and $\sigma$-s.
The output of the network are the diffuse, specular, roughness, normal parameters of an \ac{svBRDF} that, when rendered using a camera colocated flash light, look the same as the input. Our unsupervised setting allows us to fine-tune our trained network on materials to acquire.
}

Comparison, \textureComparison, of two textures is not trivial.
Pixel-by-pixel comparison is typically not suitable to evaluate visual statistical similarity.
Instead, images are mapped to a feature space in which images that are perceived as similar textures, map to similar points~\cite{portilla2000parametric}.
Different mappings are possible here.
Classic texture synthesis \cite{heeger1995pyramid} uses moments of linear multi-scale filters responses.
\citet{gatys2015texture} proposed to use Gram matrices of non-linear multi-scale filters responses such as those of the VGG \cite{simonyan2014very} detection network.
Such a characterization of textures was also used by \citet{aittala2016reflectance} and, without loss of generality, will be used and extended in this work as well.\\
While \materialModel is stationary, \flashImage is not --due to the lighting-- and has features at different random positions \location which are compared as
\begin{equation}
\textureComparisonOld(\flashImage_1,\flashImage_2)
:=
\expectation_{\location\sim(0,1)^2,\scale\sim(0,1)}[
|
\patch(\flashImage_1,\location,\scale)
-
\patch(\flashImage_2,\location,\scale)
|_1 ],
\label{eq:TextureComparison}
\end{equation}
where $\patchOld(\flashImage, \location)$ crops a patch of \new{randomly chosen scale \scale} at the location \location and resamples it to the input resolution of VGG \cite{simonyan2014very}, computes the filter responses and their Gram matrices:
\begin{equation}
\patch(\flashImage,\location,\scale):=
\mathtt{gram}(\mathtt{vgg}(\mathtt{resample}(\mathtt{crop}(\flashImage,\location, s)))).
\label{eq:Patch}
\end{equation}

Minimizing \materialModelParameters with respect to \refEq{ImageLoss} for a given \flashImage results in a material.
$\materialModel_\materialModelParameters$ can represent different approaches.
\citet{aittala2016reflectance} directly use the pixel basis and optimize discrete material maps for \materialModelParameters using a single input flash image \flashImage.
With their approach, optimizing for both visual similarity and stationarity is challenging.
In particular, the reflectance stationarity term \stationarity, requires a ``spectral preconditioning'' step as explained in their paper.
Instead, we propose an approach in the form of a neural model \materialModel that is (i) defined on the infinite domain and (ii) stationary by construction.
Thus, our loss does not need to include a stationarity term.

Next, we describe how to generate RGB textures using deep learning (\refSec{TextureSynthesis}), before combining the two components (flash images and NN texture (spaces)) into our approach (\refSec{OurApproach}).

\mysubsection
{Deep Texture Synthesis}
{TextureSynthesis}
\citet{julesz1965texture} define textures by their feature statistics across space. The choice of which features to use remains an important open problem.
With the advent of deep learning, \citet{gatys2015texture} suggested to use Gram matrices of activations of filters learned in deep convolutional neural networks (\eg VGG~\cite{simonyan2014very}), for neural style transfer. \citet{aittala2016reflectance} rely on the same statistics to recover material parameters of stationary materials. By optimizing directly over pixel values, their method can produce images with the desired texture properties.
These methods however require a different long optimization to be ran for each material.

Another group of recent methods \cite{ulyanov2016texture, johnson2016perceptual} introduce neural networks capable of producing RGB textures directly, in milliseconds.
While these approaches use a network to generate the textures, they are still limited to the input texture exemplar, and do not show further variations in their results.
\citet{ulyanov2017improved} introduced an explicit diversity term enforcing results in a batch to be different.
This diversity is however limited and restrict the results quality. Indeed, they add a diversity term to the loss, but the architecture is not modified to enable it.
Alternatively,  adversarial training has been used to capture the essence of textures \cite{shaham2019singan, bergmann2017learning}, including the non-stationary case \cite{zhou2018non} or even within a single image \cite{shaham2019singan}.
In particular, StyleGAN \cite{karras2019style} generates images with details by transforming noise using adversarial training.
As opposed to these approach we do not rely on challenging adversarial trainings, by directly learning a Neural Network to produce VGG statistics.

Instead of incentivizing stationarity in the loss, \citet{henzler2019learning} suggest a learnable texture representation that is built on mapping an infinite noise field to a field that has the statistics of the exemplar texture.
Their method is a point operation, implemented by an MLP that is fed exclusively with noise sampled at different scales as done by \citet{perlin1985noise}.
By explicitly preventing the network to access any absolute position, this approach is stationary by-design. Inspired by this approach our architecture enforces a convolutional stationary by-design constraint.

\mysection{Noise to BRDF Texture Spaces}{OurApproach}

An overview of our approach is shown in \refFig{Architecture}.
We train a neural network which acts as a decoder 
$
\materialModel_\materialModelParameters(\location|\materialCode)
$
that generalizes across spatial positions \location as well as across materials, expressed as latent material codes \materialCode.
The material codes \materialCode are produced by an encoder \flashImageEncoder with $\materialCode= \flashImageEncoder(L)$.
Both encoder and decoder are trained jointly over a set of flash images using the loss:
\begin{align}
\label{eq:Loss}
\loss(\materialModelParameters)
:=
\expectation_{\flashImage}
[
\textureComparison(
\flashImage, 
\render\materialModel_\materialModelParameters(\cdot|\flashImageEncoder_\materialModelParameters(\flashImage))
)
].
\end{align}
This equation is an adapted version of \refEq{ImageLoss} to fit our objectives.
In particular we propose a neural network-based $\materialModel_\materialModelParameters$, leveraging the expectation $\expectation_{\flashImage}$ over all flash images in our training set and removing the stationarity term as it is enforced by construction in our network architecture.
We describe the flash image encoder \flashImageEncoder (\refSec{Encoder}), the material texture decoder \materialModel (\refSec{Decoder}), the texture comparison model \textureComparison (\refSec{TextureComparison}) and our fine tuning approach (\refSec{Finetuning}), next.

\mysubsection{Encoder}{Encoder}
The encoder \flashImageEncoder maps a flash image \flashImage to a latent code \materialCode. The flash images used by our method are similar to those of recent \ac{svBRDF} acquisition papers~\cite{aittala2016reflectance, deschaintre2018single}: we use a phone with a flash collocated with the camera and capture surface in a fronto-parallel way.
Our encoder is implemented using ResNet-50 \cite{he2016deep}.
The ResNet starts at a resolution of 512$\times$ 384 and maps to a compact latent code.
Empirically, we find a $\latentDimension=$ \Zdimension-dimensional latent space to work best for our data and present all results using this number.

\mysubsection{Decoder}{Decoder}
The decoder \materialModel maps location \location, conditioned on a material code \materialCode to a set of material parameter maps.
The key idea is to provide the architecture with access to noise, as previously done for style transfer \cite{huang2017arbitrary}, generative modelling \cite{karras2019style} or 3D texturing \cite{henzler2019learning}.
In particular, we sample rectangular patches with edge length of $n\times m$ pixels from an infinite random field and convert them to material maps using a U-net architecture \cite{ronneberger2015u}.
The U-net starts at the desired output resolution $n\times m$ and reduces resolution four times using max-pooling before upsampling back to $n\times m$ through a series of bi-linear upsampling and convolutions.
Let $F$ be the array of input features.
For $i=0$, the first level, in full resolution, these features are sampled from the random field at \location.
Then, output features are \begin{align}
F':=
\mathtt{adaIN}(\mathtt{conv}_\theta(F), \mathsf T_\theta\materialCode),
\end{align}
where
$\mathtt{adaIN}$ is \ac{AdaIN} \cite{huang2017arbitrary}, 
$\mathtt{conv}$ a convolution (including up- or down-sampling and ReLU non-linearity), $z$ is a latent material code and $\mathsf T$ is an affine transformation.
Components with learned parameters are denoted with subscript $\theta$. 

We use \ac{AdaIN} as defined by \citet{huang2017arbitrary} as
\begin{align}
\mathtt{adaIN}(\bm\xi, \{\bm\mu, \bm\sigma^2\})=\frac{\bm\sigma}{\bm\sigma_F}(\bm\xi-\bm\mu_F)+\bm\mu
\end{align}
and remaps the input features with mean $\bm\mu_F$ and variance $\bm\sigma_F^2$ to a distribution with mean $\bm\mu$ and variance $\bm\sigma^2$.

The affine mapping $\mathsf T$ is implemented as  $(\latentDimension+1)\times(2\times c_i)$ matrices multiplied with the latent code $z$.
Here $2\times c_i$ represent a different mean and variance for each channel dimension $c_i$ of a layer.
It provides the link between the material code and the noise statistics.
Each material code \materialCode is mapped to a mean and variance to control how the statistics of features are shaped at every channel on every layer of the decoder.\\
Our control of noise statistics from latent codes is similar to StyleGAN \cite{karras2019style}, with the key difference that we do not sample noise at different scales, but learn how to produce noise with different, complex, characteristics at different scales by repeatedly filtering it from high resolutions.

\mysubsection{Images Comparison}{TextureComparison}

\new{As mentioned in \refSec{Flash Images} (\refEq{TextureComparison} and \refEq{Patch}), we want to evaluate visual similarity and stationarity. To this end,} we propose to compare images based on a loss that accounts both for the statistics of activations \cite{gatys2015texture} and their spectrum \cite{liu2016texture} on multiple scales across the infinite spatial field,
\begin{align}
\textureComparison(\flashImage_1,\flashImage_2)
&:=
\expectation_{\location\sim\mathbb R^2,\scale\sim(s_\mathrm{min},s_\mathrm{max})}[
|
\patch(\flashImage_1,\location,\scale)
-
\patch(\flashImage_2,\location,\scale)
|_1 ].
\\
\patch(\flashImage,\location,\scale)&:=
\mathtt{gram}(V(\flashImage,\location,\scale)) + \spectralWeight\cdot\mathtt{powerSpectrum}(V(\flashImage,\location,\scale))\\
V(\flashImage,\location,\scale)&:=\mathtt{vgg}(\mathtt{resample}(\mathtt{crop}(\flashImage,\location, s)))
\end{align}
\paragraph{Spectrum}
VGG Gram matrices capture the frequency of a feature appearance, unless it forms a regular pattern \shortcite{liu2016texture}.
\citet{liu2016texture} proposed to include the L1 norm of the power spectra of RGB images into the texture metric for texture synthesis.
We combine both ideas and use VGG, but do not limit ourselves to its Gram matrix statistics, and also leverage its spectrum. \new{We set $\lambda=1e-3$.}

\paragraph{Scale}
As VGG works at a specific scale of features it was trained for, it behaves differently at different scales.
As the material should be visually plausible regardless of its scale we include multiple scales \scale, ranging from $\scale_\mathrm{min}=0.1$ to $\scale_\mathrm{max}=8$ in the loss computation.

\paragraph{Infinity}

\mycfigure{Infinite}{
\textbf{Infinite spatial extent.} 
Top result is sampled at high resolution (256$\times$4096) from our BRDF space while the bottom result is a result from \citet{guo2020materialgan} at 256$\times$256 resolution and horizontally tiled 16 times to achieve high resolution. The absence of repetitiveness in the top result demonstrates that our learned BRDF space can be sampled at any query (x,y) location without producing visible repetition artifact. By construction, our network architecture does not require any special boundary alignment to avoid tiling artifacts. 
}
Expectation over the infinite plane is implemented by simply training with different random seeds for the noise field. This results in the generation of statistically similar, but locally different variations of materials.
As, given a seed, every generated patch is a coherent material, combinations of multiple patches remains coherent as well.
This allows to query an endless, seamless and diverse stream of patches without repetition.
It also prevents over-fitting and is crucial to guarantee stationarity by-design.

\mysubsection{Training}{Training}
\new{To enforce a generalizable material prior}, we \new{first} train the system as a \ac{VAE} \cite{kingma2013auto}. Instead of mapping to a single \Zdimension-D latent material code, the encoder \flashImageEncoder maps to a \Zdimension-D mean and variance vector, from which we sample in training.
At test time we use the mean for each \Zdimension-D.
We have omitted the additional \ac{VAE} terms enforcing \materialCode to be normally distributed from \refEq{Loss} and \refFig{Architecture} for clarity.
We trained our model for 4 days and a batch size of 4 on a NVIDIA Tesla V100 using ADAM optimizer with a learning rate of 1e-4 and weight decay 1e-5.

\mysubsection{Fine-tuning}{Finetuning}
\new{Using the trained} encoder-decoder pair we can instantaneously compress a 2D RGB flash image to a latent code and decompress into an infinite \ac{svBRDF} field.
The quality of the decoding can further be improved by adapting the decoder weights to a specific exemplar $\fineFlashImage$ with a short one-shot training.  
To this end, all weights $\materialModelParameters$ are held fixed, except for the decoder weights $\fineMaterialModelParameters\subset\materialModelParameters$, which are further trained to reproduce a single flash image $\fineFlashImage$ at material code $\fineMaterialCode=\flashImageEncoder(\fineFlashImage)$.
This is made possible by our completely unsupervised approach, allowing to fine-tune for any flash image, without requiring ground truth maps. \new{Note that unlike \cite{guo2020materialgan} we use a style loss rather than a pixel-wise loss for fine-tuning, preserving the \emph{diversity} properties of our results.}
In practice we fine-tune for 1000 steps \new{with an increased learning rate by a factor of 10}, for about 5 minutes.

Fine-tuning of two materials will result in two different decoders $\materialModel_1$ and $\materialModel_2$ as well as two latent codes $\materialCode_1$ and $\materialCode_2$ produced by the same encoder.
We show that despite being a more complex space, interpolating both the latent code and decoder parameters, as in $\mathtt{lerp}(\materialModel_1, \materialModel_2)(\mathtt{lerp}(\materialCode_1, \materialCode_1))$ works well in practice,
Unless otherwise specified, we show fine-tuned results in the remainder of this paper and ablate several variants in \refSec{Ablations}.

\mysubsection{Material model}{MaterialModel}
We use the Cook-Torrance \shortcite{cook1982reflectance} micro-facet BRDF Model, with Smith's geometric term \cite{heitz2014understanding}, Schlick's \shortcite{schlick1994inexpensive} Fresnel and GGX \cite{walter2007microfacet}.
Hence, parameters are diffuse RGB albedo, monochromatic specular albedo, roughness and height, \ie six dimensions.
Instead of learning a normal map, a height field is generated from which normals are computed using finite differences. During the our differentiable rendering step, we assume a FOV of $45^{\circ}$ to simulate smartphone cameras.

\mysubsection{Alignment}{Alignment}
Many flash images entail a slight rotation as it can be difficult to take a completely fronto-parallel image.
This was handled by \citet{aittala2016reflectance} by locating the brightest pixel and cropping, but we found our, more abstract, training to struggle with such a solution.

Instead, we add a horizontal and a vertical rotation angle to the parameter vector generated from the latent code (not shown in \refFig{Architecture} for clarity).
During training, these are used to rotate the plane, including the normals.
During testing, these angles are not applied meaning that the output is in the local space of the exemplar.

We use a branch of the encoder to perform the alignment task, allowing to jointly align images based on their visual features.

A byproduct is that the encoder returns angular distance to fronto-parallelity, which could be used to guide users during capture.

\mycfigure{ResultComparison}{
\textbf{Comparison with other methods.} Each method \textbf{(rows)} decomposes an image into \ac{svBRDF} parameters \textbf{(columns)}.
The first column shows the flash image input and the second column the rendering of the results under a similar fronto-parrallel lighting.
The third column is the material relit from the top, showing the generalization capacity across light. Our method's quality is particularly visible under a novel illumination (see also \refFig{Relighting}).
This is because other methods leave a trace of the flash in the \ac{svBRDF} maps, as can be seen in the decomposed channels (four right-most columns). \new{These results are obtained with our single image setting, compared methods \citet{gao2019deep} and \citet{guo2020materialgan} could benefit from additional aligned images or accurate light calibration when available.}
Please see the supplemental material for similar results on many more materials.
}

\mycfigure{Interpolation}{
\textbf{Interpolation of latent BRDF texture codes.}
In each row, a left and a right latent code and generator weights $\mathbf z_1$, $\mathbf z_2$, $\mathbf g_1$, $\mathbf g_2$ are obtained by encoding two flash images, respectively.
The intermediate, continuous field of BRDF parameters is computed by interpolating, in the learned BRDF space, from $\mathbf z_1$ \& $\mathbf g_1$ to $\mathbf z_2$ \& $\mathbf g_2$ and conditioning the decoder \ac{CNN} with the intermediate codes.
The result is lit with a fronto-parallel light source to demonstrate the changes in appearance. For comparison, the first row shows image space linear interpolation, the second compares to \citet{guo2020materialgan}. The third row shows an ablation of our approach trained on a single material (without previous full dataset training). This lack of training prevents it from creating a cohesive space in which to interpolate. Overall our approach allows for interpolation, progressively changing both structure and reflectance.
}

\mysection{Results}{Results}

\mysubsection{Dataset}{Dataset}
We created an extended a dataset of flash images for testing and training of our approach.
It comprises of 356 images of various types of materials we captured using \new{four different} smartphones.
We reserve 50 images for testing, augmented by all images from \citet{aittala2016reflectance}.
Hence, no image from \citet{aittala2016reflectance} was used for training.

\mysubsection{Quantitative Evaluation}{Quantitative}
For quantitative analysis we compare our approach to a range of alternative methods with respect to different metrics.

\paragraph{Methods}
We compare to five methods by (i)~\citet{aittala2016reflectance}, (ii)~\citet{deschaintre2018single},
(iii)~\citet{gao2019deep},
(iv)~\citet{guo2020materialgan},
and (v)~\citet{zhao2020joint}.
All renderings of these methods are done with the material model described in their respective paper. \new{While \citet{gao2019deep} and \citet{guo2020materialgan} were designed to be compatible with multiple image acquisition with known light positions, in our comparisons we provide the same input as to our method: a single input image and an approximate light position.}

\paragraph{Metrics}
We quantify \emph{style}, \emph{diversity}, and \emph{computational speed}.
Style is captured by L1 difference of the VGG Gram matrices of rendered images.
A good agreement in style has a low number \ie less is better. We also evaluate XYZ histogram L1 difference and find that all methods have below 1\% of difference with Ground Truth renderings, indicating good color matching for all. Histogram difference does not however capture the complex visual difference when comparing materials (as can be seen in \refFig{Relighting}).
Diversity is captured as the mean pairwise VGG L1 across all realizations \cite{henzler2019learning}.
Here, more is better.
The idea behind this diversity metric is, that a for a diverse method, two realizations should have a high difference.
A direct pixel metric would be sensitive to noise which generates small perturbations resulting in false-positive differences.
Hence, the choice of VGG features, which detects if realizations are indeed perceivably different. 
Note that we do not evaluate pixel-wise metrics such as L1 or SSIM as these enforce local coherence, which is, by construction, not targeted by our method.

\paragraph{Comparisons}

We use the described metrics to compare against multiple state-of-the-art methods in material acquisition and report the results on real (Flash) and synthetic materials (Relit) in \refTbl{Results}. For real results, we only have access to the frontal flash-illuminated material and therefore compare the picture to a rendering of each method's result also under frontal illumination.\\
This, however, does not evaluate well the appearance under novel illumination, which is a key property of \acp{svBRDF}. To validate the generalization across light directions, we acquire 30 random stationary synthetic \acp{svBRDF} from CC0 Texture and render them to simulate a frontal-flash capture setup using Mitsuba2 \new{\cite{nimier2019mitsuba}}. All methods are then run with this simulated flash image as input. We report the average of the re-lighting error, against ground truth renderings, for all methods under 10 random point light illuminations.

As shown in \refTbl{Results}, our approach is the only one to target diverse results, \ie we produce infinitely many realizations of a texture while all other approaches produce only one.
Thus, diversity (Div.) is zero for compared methods, while our approach can generate varied realizations for each material.

In terms of computational speed, \citet{aittala2016reflectance} and \citet{zhao2020joint} both require long --between 1 and 3 hours-- per-exemplar optimization to produce a stationary texture. Our approach requires around 500 ms to generate a material and a few minutes to fine-tune it to a given input. This is in the same order of speed than \citet{deschaintre2018single} for generation and \citet{gao2019deep} and \citet{guo2020materialgan} for the fine-tuning. Once fine-tuned, our method can generate new realizations and high resolutions versions of the targeted material in around 500 ms.

\mycfigure{Relighting}{
\textbf{Relighting} of different materials (rows) using material maps extracted by different methods (columns).
The first column shows the input flash image where light is fronto-parallel.
The light in all other images comes form the top.
While no reference is available for this task, it is apparent that all the methods  except ours struggle to generalize to novel light conditions. Note that 
\citet{deschaintre2018single}, 
\citet{gao2019deep} and \citet{guo2020materialgan} leave a dark residual of the flash in the material maps.
\citet{zhao2020joint} and \citet{aittala2016reflectance} \new{fare} slightly better and avoid the residual, but the  structures do not match. \new{These results are obtained with our single image setting, compared methods \citet{gao2019deep} and \citet{guo2020materialgan} could benefit from additional aligned images or accurate light calibration when available.}
}

\begin{table}
\setlength{\tabcolsep}{2pt}
\caption{We compare to recent material acquisition approaches on the L1 difference between VGG Gram matrices (VGG Style, lower is better) on both real and synthetic results as described in \refSec{Quantitative}. Additionally we evaluate each method's capacity to generate diverse realizations of a material with the mean pairwise VGG L1 across all realizations (Div, higher is better). We see that ours outperforms others on perceived similarity with the VGG style metric. Additionally, ours is the only one generating diverse material variations from a single image.}
\vspace{-0.2cm}
\label{tbl:Results}%
\centering
\begin{tabular}{r
    S[table-format=1.3]
    S[table-format=1.3]
    S[table-format=1.2]
    }
    \multicolumn{1}{c}{Method}&
    \multicolumn{2}{c}{Style err. $\downarrow$}&
    \multicolumn{1}{c}{Div. $\uparrow$}
    \\
    \cmidrule(lr){2-3}
    &
    {Flash}&
    {Relit}&
    \\
    \toprule
    \citet{aittala2016reflectance}&
    0.922&
    0.512&
    0.00
    \\
    \citet{deschaintre2018single}&
    0.943&
    0.653&
    0.00
    \\
    \citet{gao2019deep}&
    0.738&
    0.556&
    0.00
    \\
    \citet{zhao2020joint}&
    \winner{0.545}&
    0.618&
    0.00
    \\ 
    \citet{guo2020materialgan}&
    0.843&
    0.582&
    0.00
    \\ 
    Ours&
    0.597&
    \winner{0.439}&
    \winner{2.08}
    \\
    \bottomrule
\end{tabular}
\end{table}

\mysubsection{Qualitative Evaluation}{Qualitative}

\paragraph{Decomposition}
A qualitative example of our \ac{svBRDF} decomposition (Normal, Diffuse Roughness and Specular maps) and re-renderings under different lights are depicted in \refFig{ResultComparison}. Please see our supplemental material for all results decomposition and comparison. We see that our method captures best the material behaviour and does not suffer from artefact in the over-exposed area of the input image which can be seen in previous work. As our method uses materials statistics rather than direct pixel aligned image to material transformation, it is immune to such artefacts.

\paragraph{Relighting}
In \refFig{Relighting} we show qualitative rendering comparisons on real materials with illumination coming from the top. In this more challenging setting, it is clear that existing work struggle to remove the highlight from the center of the flash image, which does not affect our method. As \citet{aittala2016reflectance} reconstruct a small (representative) patch of the large input picture, their method is also immune to flash artefacts, but result in a very zoomed representation of the material. To compensate for this "zoom factor", we tile the results in each direction. We empirically found that 3 times works best for most materials.

\paragraph{Seeds}
In \refFig{Seeds} we show the variation of our results when changing the seed. The overall appearance of the material remains the same but the details (such as the rust or the leather normals and color variation) vary.
\myfigure{Seeds}{\textbf{Seeds variation}. We vary the seed for the generation of different realizations for acquired materials, while preserving its overall appearance. The zoomed-in insets all show the same region of the material, allowing to better appreciate the variations.}

Overall, we see in \refFig{ResultComparison}, \refFig{Relighting} and \refFig{Seeds} that our approach can capture a large range of different stationary materials, reproducing their style, yet being diverse. This enables different properties described next.

\paragraph{Infinite}
We show in \refFig{Infinite} the "infinite" resolution capacity of our approach against the common approach of tiling. Our result (top image) shows no sign of repetitiveness even for very large resolution (4096$\times$256). 

\paragraph{Interpolation}
We show results of interpolation between materials, as described in \refSec{Finetuning}, in \refFig{Interpolation} and Supplemental Material. We compare against the linear interpolation baseline and \citet{guo2020materialgan} which also allows interpolation. \new{We find our method to provide smoother interpolation than the Linear approach and to better preserve intermediate material color than \citet{guo2020materialgan}}. We additionally evaluate interpolation if we directly train on material individually (without the training step described in \refSec{Training}). This confirms that this pre-training forms a coherent latent space in which we can navigate.

\paragraph{Texturing}
\refFig{Teaser} shows examples of applying maps produced by our approach to a complex 3D shape.
Thanks to our generative model, we can easily texture many sneakers, without spatial or material repetition.
At any point, a user can randomize the generated material, generate new materials from pictures or interpolate between new materials and old ones.

\paragraph{Generation}
Our \materialCode space can be sampled to generate new materials as shown in in \refFig{Generation}  with a variety of examples

\myfigure{Generation}{\textbf{Generation}. Random samples from our space. We generate new materials by sampling the \materialCode space and render them with a frontal flash. See supplemental materials for more generated materials.}

\paragraph{Interactive demo}
The visual quality is best inspected from our interactive WebGL demo in the supplemental material.
It allows exploring the space by relighting, changing the random seed and visualizing individual BRDF model channels and their combinations.
The same package contains all channels of all materials as images as well as compared methods. See the accompanying video for a demonstration of our interactive interface.

\newcommand{\ablation}[1]{\textsc{#1}}
\begin{wraptable}{r}{3.4cm}
    \centering%
    \caption{VGG style error for ablations relative to our \ablation{Full}.
    For reference, our full method has an absolute score of 0.44.}%
    \setlength{\tabcolsep}{3pt}
    \vspace*{-0.3cm}
    \begin{tabular}{r
    S[table-format=2.2,retain-explicit-plus=true]}
        Ablation&
        {Error $\downarrow$}\\
        \toprule
        \ablation{Single}&-0.5{\%}\\
        \ablation{NonTuned}&+24.0{\%}\\
        \ablation{DecoderOnly}&+2.0{\%}\\
        \ablation{Fourier}&+0.9{\%}\\
        \ablation{Light}&+1.7{\%}\\
    \bottomrule
    \end{tabular}
    \label{tbl:Ablations}
\end{wraptable}

\paragraph{Fine-tuning}
We show the results quality improvement when using the proposed fine-tuning approach in \refFig{Finetuning}. We can see that the structure and details better match the input picture.

\myfigure{Finetuning}{
\textbf{Fine-tuning}. We show results on two results of real materials reproduced using our pre-trained network (ours non-tuned) and the same material using our fine-tuning approach. We can see that our fine-tuned results match the input material appearance significantly better. Note that fine-tuning is only with image supervision and does not have access to any underlying BRDF supervision. }

\mysubsection{Ablation Experiments}{Ablations}
We study several variants of our approach to evaluate the relevance of individual contributions to our \ablation{Full} method. 

We report the results of these evaluation in \refTbl{Ablations} with VGG Style error in \refSec{Quantitative}. We did not find the diversity of our method to be affected by these ablations.

\ablation{Single} describes our method trained on a single example, without the previous training step. The results are slightly better than our \ablation{Full} method, but requires twice longer per material training and does not generalize to a space, preventing interpolation and generation of materials.

\ablation{NonTuned} is our method without the fine-tuning step from \refSec{Finetuning}, confirming that it significantly improves the match to the acquired material.
\ablation{DecoderOnly} describes the change of our generator to a decoder only architecture. We show that removing the encoder part of the generator slightly degrades the results.
\setlength{\columnsep}{8pt}
\ablation{Fourier} and \ablation{Light} respectively result from the removal of the Fourier component \new{(power spectrum)} of our loss (\refSec{TextureComparison}) and the removal of the light alignment branch of our encoder (described in \refSec{Alignment}), which both lead to slightly worse results.

\mysection{User experiment}{UserStudy}

We perform a user study to better understand the capabilities of different methods.
Our main aim is to provide material maps that robustly generalize to all light conditions so they can be deployed in production rendering.
Hence we study a \emph{relighting task}: given a the input image in one light condition, we ask humans to pick the method that looks most plausible ``in a different light''.

\paragraph{Methods}
Subjects anonymously completed an online form without time limit. 
At the start of the user study, participants were shown two photos of a marble material taken under two different lighting conditions to exemplify what a valid relighting could look like. 
They performed 10 trials, each corresponding to one material.
In each trial, they were presented a reference image rendered in one light conditions (``flash'') and six relit images in another light condition (``top'').
Relit images were displayed in a randomized spatial 2D layout.
We consider six different methods: \citet{aittala2016reflectance,deschaintre2018single,gao2019deep,guo2020materialgan,zhao2020joint} and ours.
Samples of those stimuli are seen in \refFig{Relighting}.
Participants were asked to pick the image (images were not named) that, according to them, was the best faithful relighting of the source (flash) image.
Note that no relit reference was shown. 

\paragraph{Analysis}
\setlength{\columnsep}{8pt}
\begin{wraptable}{r}{4.5cm}
    \centering
    \caption{User preferences per method.}
    \label{tbl:Study}
    \begin{tabular}{rr}
        \multicolumn1c{Method}&
        \multicolumn1c{Freq.}\\
        \toprule
        \citet{aittala2016reflectance}&21\\
        \citet{deschaintre2018single}&10\\
        \citet{gao2019deep}&4\\
        \citet{guo2020materialgan}&11\\
        \citet{zhao2020joint}&30\\
        Ours&314\\ 
        \bottomrule
        \end{tabular}
\end{wraptable}
A total of \new{$N=39$} participants completed the experiment as summarized in \refTbl{Study}.
A $\chi^2$ test rejects ($p<0.0001$) the hypothesis that choices were random.
Pairwise binomial post-hoc tests further show that our method is different from any other method, at the same significance level.
Most importantly, subjects choose our method in \new{314} out of \new{390} total answers \new{80.5}\,\%).
We did not analyze the relation of other methods relative to each other.

\mysection{Limitations}{Limitations}
 Our method relies on fronto-parallel flash acquisition. While we propose a mitigation solution in \refSec{Alignment}, we show in \refFig{Limitations} that we are not completely invariant to large light and plane rotations. Our approach is also limited to stationary isotropic materials and relies on the planarity of the captured surface. 

\myfigure{Limitations}{
\textbf{Flash acquisition assumption}. We show example of how our results degrade when fronto-parrallel, collocated flash assumptions are broken. The recovered material appearance varies (roughness, high frequency normal) but maintains the overall appearance of the input image. }

\mysection{Conclusion}{Conclusion}
We have presented an approach to generate a space of BRDF textures using a small set of flash images in an unsupervised way.
Comparing this approach to the literature shows competitive metrics for re-renderings with the unique advantage of being able to generate an infinite and diverse field of BRDF parameters.

In the future, it would be interesting to increase the complexity of supported material whether in term of shading or non stationarity. Also, not relying on fine-tuning to increase the network expressiveness would allow to create an even more cohesive space. 
Further, more refined differentiable rendering  material models could be used to derive stochastic textures, including shadows, displacement, or scattering as well as volumetric or time-varying textures.
We believe that our framework will represent a stepping stone for more complex infinite and diverse BRDFs acquisition.

\section*{Acknowledgement}
\new{This work was supported by the ERC Starting Grant SmartGeometry, a Google AR/VR Research Award, Dr. Abhijeet Ghosh and his EPSRC Early Career Fellowship (EP/N006259/1) and a GPU donation by NVIDIA Corporation. We thank the authors of \citet{aittala2016reflectance}, \citet{gao2019deep}, \citet{guo2020materialgan}, \citet{zhao2020joint} for providing their implementations and helping with the comparisons.}

\bibliographystyle{ACM-Reference-Format}
\bibliography{paper}
\end{document}